# HYPER-VISCOELASTIC 3D RESPONSE OF AXONS SUBJECTED TO REPEATED TENSILE LOADS IN BRAIN WHITE MATTER

**Mohit Agarwal and Assimina A. Pelegri***

Mechanical and Aerospace Engineering
Rutgers, The State University of New Jersey
Piscataway, NJ, USA

*Corresponding author - pelegri@rutgers.edu

## ABSTRACT

*A novel finite element method (FEM) is developed to study mechanical response of axons embedded in extra cellular matrix (ECM) when subjected to harmonic uniaxial stretch under purely non-affine kinematic boundary conditions. The proposed modeling approach combines hyper-elastic (such as Ogden model) and time/frequency domain viscoelastic constitutive models to evaluate the effect of parametrically varying oligodendrocyte-axon tethering under harmonic stretch at 50Hz. A hybrid hyper-viscoelastic material (HVE) model enabled the analysis of repeated uniaxial load on stress propagation and damage accumulation in white matter.*

*In the proposed FEM, oligodendrocyte connections to axons are depicted via a spring-dashpot model. This tethering technique facilitates contact definition at various locations, parameterizes connection points and varies stiffness of connection hubs. Results from a home-grown FE submodel configuration of a single oligodendrocyte tethered to axons at various locations are presented. Root mean square deviation (RMSD) are computed between stress-strain plots to depict trends in mechanical response. Steady-state dynamic (SSD) simulations show stress relaxation in axons. Gradual axonal softening under repetitive loads is illustrated employing Prony series - HVE models. Representative von-Mises stress plots indicate that undulated axons experience bending stresses along their tortuous path, suggesting greater susceptibility to damage accumulation and fatigue failure due to repeated strains.*

## NOMENCLATURE

| | |
|---|---|
| $\alpha$ | alpha |
| $\mu$ | shear moduli (hyper-elastic: Ogden model) |
| $\lambda$ | principal stretches |
| $\sigma$ | principal stress |
| $G$ | complex shear modulus (viscoelastic model) |
| $\tau$ | shear stress |
| $K$ | spring-dashpot stiffness values |

## 1. INTRODUCTION

Traumatic brain injury (TBI) is described as an acquired insult to the brain due to an external mechanical force that could lead to temporary or permanent impairment [1, 2]. TBI is a major health concern in the USA and around the globe and has been reported as leading cause of death and disability among children and young adults in the United states [3, 4]. An excessive mechanical loading that might occur during vehicle accidents, sports injuries, violence, or injuries related to everyday activities (e.g., impact with furniture or falling downstairs) can be the prodrome of a mild (mTBI) or a regular brain injury. These external assaults could be singular/instantaneous or repetitive in nature. Depending on intensity, moderate to severe TBIs can have long lasting or permanent effects such as cerebrovascular damage, neuronal deformation, hypoxia, cerebral edema, and increased intracranial pressure [4, 5].

Brainstem and corpus callosum are usually susceptible to TBI, as they are often subjected to high strains and at risk of severe axonal damage [6, 7]. Brain soft tissues show significant variations in overall material stiffness and critical regions exhibit highly anisotropic material properties, contributing to amplified local deformations at vulnerable sites. Thus, finite element methods (FEM) have emerged as promising tool to model, characterize stress-strain response of the brain tissue and predict response to traumatic loads of both singular and repetitive nature[8].

In the past decade, several axonal material models have been put forward including linear elastic, viscoelastic [9, 10], and hyper-elastic, to define properties for brain soft matter [11-13]. Hybrid constitutive material models combining hyper-elastic and viscoelastic properties to depict hyper-viscoelasticity (HVE) have also been proposed [14, 15]. Such model could characterize material non-linearities and time dependent strain accumulation behavior when soft biological tissues are subjected to creep, cyclic load, and transient loading scenarios [2, 16-18]. Lack of accurate material properties depicting individual axons and ECMs is still one of the biggest hurdles in realizing high fidelity brain tissue FEM [19]. Over the years, hybrid techniques such as inverse finite element analysis to predict axonal material properties have also been tried out to obtain material parameters to closest approximations [12, 20, 21]. Significant research efforts are underway in identifying optimal FE model geometry [22] and interfacing parameters between the axons and ECM to depict stress transfer under singular or repetitive traumatic loads [23]. Choosing appropriate axon-ECM kinematics is critical in depicting stress response in soft tissues. Some pioneering research in the field assumed purely affine boundary conditions to model the axons and ECMs (axons entirely tied down to the ECM). However, this approximation neglects the axonal tortuosity transitional behavior induced by the stretch/strain of the axon-glia system [2, 23].

The central nervous system (CNS) of the brain comprises of white and gray matter. White matter includes myelin coated axons and oligodendrocytes. Axons are long slender projections
of neuron which relays information to other neurons, muscles, and glands [7]. Oligodendrocytes are glial cells supporting and insulating axons via sheath of myelin, which improves axonal stiffness. In this study, a previously published proof-of-concept FEM [2, 7] has been revisited to investigate effect of oligodendrocyte tethering on stiffness response in an axon-ECM micromechanical model when subjected to repeated uniaxial stretch. The current model deploys purely non-affine boundary conditions. HVE material model featuring viscoelastic properties in both frequency and time domain are examined to understand stress response in axons.

## 2. MATERIALS AND METHODS

### 2.1. Micromechanical Finite Element Model

The microscale FEMs are developed with the aid of Abaqus 2020 and Python scripting. Representative Elemental Volume (REV) for modeling axons tethered to glia in CNS white matter is derived from FEM developed by Pan et al. [12] and further revised by Agarwal et al. [7]. Axons with varying undulations and radii are embedded in a 3D rectangular ECM with dimensions: x = 0.9 μm, y = 8 μm, z = 5.747 μm. Axonal undulation variations based on the work by Bain et al. [24], with average undulation varying from 1.00 to 1.10. In the current FEM, axonal diameters vary from minimum of 0.4 μm to a maximum of 0.62 μm with an average axonal diameter of 0.45 μm. Overall, same FEM geometrical setup used in our previously published work (see [7] for more details on FEM setup). Non-affine boundary conditions between the axons and ECM are attained via "surface to surface" contact definition (see **Figure 1**) [25].

In this paper, our model analyzes effect of oligodendrocyte tethering by simulating ensemble of connection scenarios for repetitive uniaxial stretch. As a proof-of-concept, single-oligodendrocyte (single-OL) FE submodel configuration is explored. For different connection configurations, varying stretch values are applied, and cumulative stiffness response is observed. *The intent is to understand whether HVE material model duly captures the damage accumulation in brain soft tissues for implicit repeated stretch when analyzed in both frequency and time (Prony series) domains.*

Previously investigated hyper-elastic material model parameters are employed to incorporate time and frequency domain viscoelastic parameters to define HVE material model. The viscoelastic parameters include the real and imaginary parts of $\omega_{g^*}$ and $\omega_{k^*}$ where $\omega$ (Omega) is the circular frequency (defined in cycles per time). The frequency domain parameters are enlisted in **Table 2** and described in detail in **§ Section 2.3**. These frequency domain viscoelastic parameters were derived from research by Wu et al. [9, 10]. Refer Abaqus manual for constitutive relationships between defined model parameters and frequency dependent storage and loss moduli: $G_s(\omega)$, $G_l(\omega)$, $K_s(\omega)$ and $K_l(\omega)$ [25]. The Prony series parameters for time domain viscoelasticity $g_i\ Prony$, $k_i\ Prony$ and $\tau_i\ Prony$ were calculated based on the research from Karami et al. [19]. The $k_i\ Prony$ values were derived from the user defined instantaneous $G_0$, $E_0$ and $\nu_0$ values. All Prony parameters are listed in **Table 3** and described in **§ Section 2.3**

In this study the intent is to study impact of repetitive / harmonic excitation (at fixed frequency – 50 Hz) on axonal stiffness and secondly to understand how cumulative stretch and strain-rates affects HVE defined axon's mechanical response. To achieve these objectives two analysis types in Abaqus 2020 were implemented: 1.) "direct" Steady State Dynamic (SSD) – in Abaqus/ Standard solver 2.) Explicit Dynamics (ED) solver in Abaqus Explicit STEP module. The motivation behind ED analysis is to depict influence of time-dependent viscoelastic component in HVE material model for highly non-linear deformation FEM and understand consequent stress accumulation characteristics over specified time period.

In Abaqus, direct SSD analysis helps predict the steady-state dynamic linearized response of a system to harmonic excitation/ repetitive load [25]. SSD is more accurate in computing dynamic response where viscoelastic (VE) material behavior is specified. In this paper, SSD simulations with HVE and Hyper-elastic (HE) models are conducted to analyze impact of HVE definition in capturing repetitive load related non-linear strain accumulations. From these set of simulations, stress relaxation trends in frequency domain HVE model is observed. For the time-domain HVE, amplified stiffness trends are due to the strain-rate effect on the viscoelastic material component. For the ED model, it is noted that viscoelastic stress component induced increased with the strain rates. In depth analysis of variations in mechanical response for different material and solution types are analyzed in **§ Section 3**.

Due to lack of any published literature sources characterizing the oligodendrocyte stiffness, the same spring-dashpot approximation [7] has been deployed to model the arms of the oligodendrocyte that tether to the axons. Since oligodendrocytes tether to axons via a sheath of myelin, the material properties of myelin served as the upper limit for parameterization of oligodendrocyte stiffness. Also, the same distributed coupling constraints are used for modeling the nucleus as implemented in our previous model [2, 7].

Here, the oligodendrocyte soma is depicted as a sphere of 0.025 μm embedded in the ECM. The nucleus is modeled as *distributed coupling* constraint in Abaqus. The reference node of this distributed coupling is positioned at the center of the sphere. The nodes of the ECM along the surface of the sphere act as *coupling nodes*. Typically, a *distributed coupling* constrains the motion of "*coupling nodes*" to the translation and rotation of the reference node. Such a constraint allows for the distribution of loads through a weighting factor between the reference and the coupling nodes based on a user specified radius of influence [2]. For the interested reader more info on the coupling constraints and contact definition can be found in our previous work [2, 7]. Finally, a linear spring-dashpot connects this remote point on the axon to the center of the oligodendrocyte sphere as shown in **Figure 2(b)**.

## 2.2. Hyper-elastic (HE) Material Model Component

Nonlinear hyper-elastic models are often used for simulation of soft biological tissues [8, 12, 20, 21, 26]. In this research, the Ogden hyper-elastic (HE) material model is used to simulate the ECM and the axons [2] because its non-linear response allows for more accurate characterization of the neural tissue at large deformations and strains while capturing the rate dependent behavior. The Ogden hyper-elastic model is based on the three principal stretches $\lambda_1$, $\lambda_2$, $\lambda_3$ and *2N* material constants. The strain energy density function, *W*, for the Ogden material model (Equation 1) in Abaqus is formulated as [2, 25]:

$$W = \sum_{1}^{N} \frac{2\mu_i}{\alpha_i^2}\left(\lambda_1^{-\alpha_i} + \lambda_2^{-\alpha_i} + \lambda_3^{-\alpha_i} - 3\right) + \sum_{1}^{N} \frac{1}{D_i}(J_{el} - 1)^{2i} \quad (1)$$

where $\bar{\lambda}_i = J^{-\frac{1}{3}} \lambda_i$ and $\overline{\lambda_1 \lambda_2 \lambda_3} = 1$. Here, $J$ represents local change of volume and is related to the determinant of the deformation gradient tensor $F$, via the right Cauchy-Green tensor ($C = F^T F$) as $J^2 = det(F)^2$, $\mu_i$ represents shear moduli, while $\alpha_i$ and $D_i$ are material parameters. In Equation 1, $J_{el}$ is the elastic volume ratio. The first and the second terms represent the deviatoric and hydrostatic components of the strain energy function. The parameter $D_i = \frac{2}{K_0}$, allows for the inclusion of compressibility where $K_0$ is the initial bulk modulus. The same single parameter Ogden hyper-elastic material is used in this study as well [7]. Therefore, N = 1. Incompressibility implies that $J_{el} = 1$ and is specified in Abaqus by setting $D_1 = 0$. As a result, Abaqus eliminates the hydrostatic component of the strain energy density equation, and the expression reduces to the following

$$W = \sum_{1}^{N} \frac{2\mu_i}{\alpha_i^2} (\lambda_1^{-\alpha_i} + \lambda_2^{-\alpha_i} + \lambda_3^{-\alpha_i} - 3) \qquad (2)$$

For the Ogden HE model, three principal Cauchy stresses values are derived by differentiating $W$ with respect to the extension $\lambda$. For incompressible material under uniaxial tension ($\sigma_y = \sigma_z = 0$). Since the current FEM is based on uniaxial tension, the corresponding hyper-elastic constitutive model principal stress $\sigma_{uniaxial}$, can then be expressed as:

$$\sigma_{uniaxial} = \frac{2\mu}{\alpha} [\lambda^\alpha - \left(\frac{1}{\sqrt{\lambda}}\right)^\alpha] \qquad (3)$$

Undulation prevents axons from experiencing full tension until a threshold strain is attained and then undulation for the axon becomes 1. In this paper, the values for shear modulus for the axons and ECM are derived from research by Wu et al. [9] while $\alpha$ is based on the model developed by Meaney [26]. The shear modulus of the ECM is assigned relative to the shear modulus of the axon, considering axons are three times stiffer than ECM as reported by Arborgast and Marguile's published work [27]. The same methodology has been deployed to model incompressibility for the HE material modeling component [2, 7].

**Table 1** summarizes the material properties, $\mu$, $D$, and $a$, and element definition used in the steady state dynamic HE FE model.

### 2.3. Viscoelastic (VE) Material Model Component

In this study, a viscoelastic constitutive component helped define the micro-scale representative volume element (RVE) characteristics to enable time-dependent analysis of soft-tissues. The viscoelastic constitutive material model parameters in frequency domain are based on research by Wu et al. [9, 28, 29] and Sullivan et al. [10, 29]. The frequency domain viscoelastic data is obtained from both tensile and pure shear tests performed on the RVEs with axonal volume fractions (VFs) in the range of 5%-85% [9], are presented in **Table 2**.

In order to characterize mechanical response of axons in brain white matter under a repetitive impact loading scenario, time-domain viscoelastic properties has been incorporated [14, 30]. While several models have been proposed to represent brain tissue behavior over varying conditions, linear viscoelastic material model was chosen for the proposed FEM [15, 19]. Among several constitutive material models proposed to characterize mechanical behavior of brain tissues under different conditions, linear viscoelastic material model has shown the best agreement with experimental results under smaller deformation scenarios [31]. Hence, in this paper we have followed the same time-domain viscoelastic model as followed by Karami et al. [19] to formulate the Prony series parameters for time-domain HVE material modelling. In linear viscoelastic model, there is a linear relationship between strain history with respect to current stress value, see Equations (4) and (5):

$$d\varepsilon(t) = J(t - \tau) \frac{d\sigma(\tau)}{d\tau} d\tau \qquad (4)$$

$$\varepsilon(t) = \int_0^t J(t - \tau) \frac{d\sigma(\tau)}{d\tau} d\tau \qquad (5)$$

The above equation assumes that the stress $\sigma(\tau)$, is continuous and differentiable in time, and is related to the strain, $\varepsilon(t)$, via the creep function $J$. In Equation 5, $\varepsilon(t)$ denotes the complete strain history at a time $t$, obtained by integrating the strain increments from time $0$ to time $t$, over all the increments $d\tau$. In the above Equations (4) and (5), $\tau$ denote the instantaneous time parameter [32]. Boltzmann postulated that change of strain $d\varepsilon$, which depends on the complete stress history up to time $t$, would be related to the increment of stress $d\sigma$ at the specific time increment from $\tau$ to $t$ through the creep function $J$ at the time $(t - \tau)$ as shown in Equation (4). The complete strain at time $t$ can be obtained by integrating the strain increments, over $d\tau$, as shown in Equation (5). Similarly, the increment of stress over time can be described as function of strain increments and stress relaxation function $G$ over time $(t - \tau)$. The stress can be expressed over entire strain history [32], Equation (6):

$$\sigma(t) = \int_0^t S(t - \tau) \frac{d\varepsilon(\tau)}{d\tau} d\tau \qquad (6)$$

The above viscoelastic constitutive relationship can be generalized in 3D in tensor form for a given strain history and stress relaxation tensor function, $S_{ijkl}$. In the proposed FEM, uniaxial stretches are applied at time $t = 0$. The strain history and stress constitutive relationships can be described as Equations (7) and (8):

$$\varepsilon_{ij}(t) = \int_0^t J_{ijkl}(t - \tau) \frac{d\sigma_{kl}(\tau)}{d\tau} d\tau \qquad (7)$$

$$\sigma_{ij}(t) = \int_0^t S_{ijkl}(t - \tau) \frac{d\varepsilon_{kl}(\tau)}{d\tau} d\tau \qquad (8)$$

Here indices, $i, j, k, l = 1,2,3$. In Equation (7), $J_{ijkl}(t - \tau)$ represent the components of the tensorial creep function, $J(t)$, and $S_{ijkl}(t - \tau)$ in Equation (8) denote the components of the tensorial stress relaxation function, $S(t)$. The time dependent coefficient matrices $J_{ijkl}(\tau)$ in Equation 7 represents a 6 x 6 symmetric compliance coefficients matrix for the time-domain linear viscoelastic model. This simplification stems from the linear relationship between the strain history and current stress value at time instance, $\tau$ [19].

For the proposed FEM, shear $G(t)$ and bulk moduli $K(t)$ are expressed in time domain using Prony series parameters. They are expressed as shown in Equations (9) and (10), respectively.

$$G(t) = G_0 \left(1 - \sum_{i=1}^n g_i \left(1 - e^{-t/\tau_i}\right)\right) \qquad (9)$$

$$K(t) = K_0 \left(1 - \sum_{i=1}^n k_i \left(1 - e^{-t/\tau_i}\right)\right) \qquad (10)$$

where, $G_0$ and $K_0$ are the instantaneous shear and bulk moduli and $g_i$, $k_i$ and $\tau_i$ are material-dependent coefficients. In the current paper, we assume time-domain linear viscoelastic constitutive behavior in Prony series form with two terms for axons and ECM. The Prony series parameters defined in the FEM are enlisted in **Table 3**.

## 2.4. Hyper-Viscoelastic (HVE) Material Model

Considering the time-dependent nature of repetitive uniaxial loads in the current study, it was surmised that brain white matter could be best computationally modeled as hyper-viscoelastic (HVE) in nature. Rheologically HVE model involves an equilibrium spring in parallel with a Maxwell element as described in research by Li et al. [33]. The Maxwell element has damper in series with an intermediate spring. For HVE, the total Cauchy stress ($\sigma_T$) can be estimated as sum of the stress caused by the equilibrium ($\sigma_{eq}$) and intermediate spring ($\sigma_{ov}$), respectively. The viscous damper allows the force in the intermediate spring to vary with strain rates. The springs are considered hyper-elastic (HE) and the viscous damper is controlled by stain rate parameter to account for time-dependent non-linear strain responses in this hybrid material model. Range of applied strain rates plays a detrimental role in resultant HVE behavior. If the rate is too high, the damper cannot react to deform adequately and thus both springs compress equally. This leads to overstress in the rheological model. On the other hand, if the strain rate is extremely low, then all the stress is contributed by the equilibrium spring. Thus, from rheological modelling standpoint, Equation (11), describes the formula to compute total Cauchy Stress ($\sigma_T$) of the Hyper-viscoelastic (HVE) material model :

$$\sigma_T = \sigma_{eq} + \sigma_{ov} \qquad (11)$$

For quasi-linear viscoelasticity, stress relaxation function $S(t,\varepsilon)$ could be divided into strain-dependent $\sigma^E(\varepsilon)$ and time-dependent parts $s(t)$. The strain-dependent component $\sigma^E(\varepsilon)$ can be obtained from Ogden strain energy density equation (1), while the time dependent component $s(t)$ can be derived as a summation of Prony Series exponential relaxation functions, *cf.* Equations (12) and (13). Combining these characteristics, the total stress $\sigma(t)$, in the integral form can be described as a function of both time and deformation components, *cf.* Equation (14).

$$S(t,\varepsilon) = s(t)\sigma^E(\varepsilon) \qquad (12)$$

$$s(t) = \sum_{i=1}^{n} S_i e^{-\beta_i t} \qquad (13)$$

$$\sigma(t) = \int_o^t \sum_{i=1}^{n} S_i e^{-\beta_i(t-\tau)} \left[\frac{d\sigma^E}{d\varepsilon}\frac{d\varepsilon}{d\tau}\right] d\tau \qquad (14)$$

In the above stress equation formulation, $\beta_i$ values correspond to the applied strain rates. The shear coefficient ($S_i$) values are normalized and constrained such that their sum was less than unity. Details about constitutive relationships and direct FE formulation for HVE material model in Abaqus [25] have been summarized in research by Hajhashemkhani et al. [14]. Somarathna et al. [30] also provided a mathematical model on rate-dependent Ogden strain-energy potential and the true Cauchy stress computation equation to account for viscoelastic strain-rate effects. In Abaqus 2020, at material definition step, the parameters summarized in **Tables 2** and **3** were used to model the HVE characteristics to study the rate-dependent stress-strain behavior in the axons brain white matter.

## 2.4. Finite Element Submodel

In the current study, one of our in-house FE submodel (single-OL FEM / Submodel-2) presented in previous work [2, 7] is used as test setup to study the effect of oligodendrocyte tethering on mechanical response of axons for HVE material model under repeated load scenario. The single-OL FEM is evaluated for both frequency and time domain HVE material definition to plot axons stiffness response variation.

In this FE submodel (called as single-oligodendrocyte case), a single oligodendrocyte connects to all the nearby axons embedded at different sites. The single oligodendrocyte is placed at the center of the ECM. Here, tethering between axons and oligodendrocyte are parameterized to execute an ensemble of simulation cases and gain perspective on mechanical response for each connection configuration (see **Figure 3**). As mentioned before, spring-dashpot elements simulate the tethering arms of the oligodendrocyte in proposed FEM [7]. To the best knowledge of the authors, no published literature sources are available which could characterize oligodendrocyte stiffness accurately. Hence, stress-strain response of the axons was obtained by parametrically changing spring-dashpot connection ('$K$'). The myelin material properties served as the upper limit during definition of oligodendrocytes stiffness.

**FE model Boundary Conditions**: Symmetric boundary conditions applied at the top and bottom faces in x-coordinate direction and side faces in y-coordinate direction. Constraints are applied in the z-direction using fixed boundary conditions (B.C.) on one face and a repeated stretch is applied to the opposite face using a non-zero displacement boundary conditions. The repeated uniaxial stretch is defined via an amplitude curve in Abaqus CAE. In the FE setup, steady-state dynamic (SSD) step type is defined for both frequency domain and time domain HVE analysis. To depict, time dependent strain accumulation affects, explicit dynamics (ED) step type was also executed on the single-OL FEM. An implicit time integration solver technique is used in Abaqus for computation. Contact stabilization helps prevents rigid body modes before contact is made between interacting axons and ECM surfaces [5].

## 3. RESULTS AND DISCUSSION

For the present analysis, the developed micromechanical FE model is subjected to uniaxial repeated stretch in the *z*-direction. As a representative case, FEM contour plots results of single-OL FEM with 5-nodes per axon and ECM material combination have been presented for SSD and ED FE cases. The shear moduli for axon and ECM derived as discussed in §2 (using **Tables 2** and **3**). Steady-state dynamic FE simulation step was defined (**Figure 4**). For the shown contour plots, time-domain viscoelastic properties (Prony series) were employed in for the HVE axon and ECM material (**Figures 4** and **5**). It is observed that for 20% stretch in the *z*-direction similar stress contour plots were obtained (as in [2]) and tortuosity again prevented full extensions in axons (see **Figure 5**) [34].

Bending stresses in axons over stretch history increases risk of fatigue failure. Since HVE -ED model captures non-linear strain history characteristics, von-Misses stress values are observed to be approximately 62% higher ED FEM compared to SSD FE setup for same configuration (see **Figure 5**). ED FEM results does account cumulative strains over time. Thus, exhaustive FE setup could help setting up fatigue damage model in future [2].

### 3.1. Single-OL FEM – HVE in Frequency domain

The baseline model proposed here, studies the direct steady-state dynamic (SSD) analysis implemented between a Hyperelastic (HE) versus a Hyper-viscoelastic (HVE) material models. In Abaqus, SSD step module is used to calculate SSD [25] linearized response of a system when subjected to harmonic excitation at a given freque 50Hz. SSD model definition is particularly useful when viscoelastic material behavior is present in the model. Frequency spacing kept default logarithmic scale [25] for this specific case.

The RMSD between the HE and HVE material model curves showed distinct softening of axons in HVE model, i.e., root mean square deviation (RMSD) between the 45 connections (5-nodes per axon – HE) and 45 connections (5-nodes per axon – HVE material property) curves is 0.004243 at 100*K*. Please Note: Here, 100*K* is simply denoting oligodendrocyte spring stiffness value of 100 N/m as 100K. Similar nomenclature is followed for 10 N/m as 10*K*, 50 N/m as 50*K* and 75 N/m as 75*K* respectively [7] to describe oligodendrocyte spring stiffness parameter. Root-Mean-Square Deviation (RMSD) is defined as $\sqrt{\sum \frac{(f(x_i) - g(x_i))^2}{N}}$ for curves $f(x)$, $g(x)$ and $N$ being number of points $x_i$ at which curves are compared[7]. The RMSD between the curves at 10K was found to be 0.004388. As shown in **Figure 6**, with increasing '$K$' greater decrease in axonal stiffness noticed. While RMSD between HE and HVE for each corresponding '$K$' value were observed to be similar i.e. ( RMSD for 10K case : 0.004388 > 100K case : 0.004242). Similar trend seen in 2-nodes per axon (18 connections) single-OL FEM. This decrease in axonal stiffness could be attributed to the stress-relaxation characteristics of the viscoelastic component in HVE. The dissipative part of the material behavior is governed by tabulated viscoelastic parameters for the real and imaginary parts of $g^*$ and $k^*$ (as functions of frequency) as shown in **Table 2**.

## 3.2. Single-OL FEM – HVE in Prony Series:

Using the parameters from §2.3, the viscoelastic material component in HVE was changed from frequency domain to time-domain (Prony-series) [15]. The intent is to observe time dependent strain effects on the axonal stiffness. In Abaqus, SSD step module [25] is used for HVE – Prony series (HVEPS) FEM analysis setup. Prony series parameters defined as derived in **Table 3** (§2.3).

As discussed in §2.4, the stress relaxation function has time-dependent and strain-dependent components. The $\sigma^E(\varepsilon)$ component which is obtained from Ogden strain density function increases with greater $K$ values as observed in our purely HE models in [2, 7]. The time dependent component $s(t)$, is dependent on the strain rate $\beta$ (negative exponential function).

To fit purely hyper-elastic (HE) material component coefficients in HVE model, the strain rates are considered quasi-static, to negate damper affects in the Maxwell component. In this research, we already obtained HE Ogden parameters from previous research [7, 22]. Hence, for chosen quasi-linear viscoelastic HVE model, at relatively low strain rates, i.e. when $\beta$ term is less than 1 [15]. The time-dependent $s(t)$ function gets magnified and as stretch values are increased from 10% to 100% at the same excitation frequency, greater increase in overall $S(t,\varepsilon)$ factor is observed. Since HVEPS defined SSD analysis takes into the viscous damper (strain-rate dependent) contributions. Thus, greater stiffness is observed in HVEPS when compared against frequency domain HVE model of axons and ECM, *cf.* **Figure 7**.

### 3.2.1. For Varying "K" values:

For single-OL FEM – SSD model, the axonal stiffness response was evaluated for varying $K$ = 10, 50, 75 and 100 N/m.
In the single-OL FEM model, two connection cases were evaluated 2-nodes per axon (18 connections) and 5-nodes per axon (45 connections) for varying $K$ values. Refer to **Figure 8 (a)** and **(b)**. For both tethering scenarios, axonal stiffness response decreased with increasing $K$ value. For 45 connection case in **Figure 8 (a):** 10$K$ and 50$K$ (0.001362) < RMSD between 10$K$ and 75$K$ curves (0.00215) < RMSD between 10$K$ and 100$K$ (0.00289). Similar trend was noticed for 18-connections model **(b)**, RMSD between 10$K$ and 50$K$ (0.00099) < RMSD between 10$K$ and 75$K$ curves (0.0016) < RMSD between 10$K$ and 100$K$ (0.00214).

As described in §2.4 for HVE rheological models that strain-rate dependent behavior cannot be observed if the system is subjected to either quasi-static (extremely low strain rate) or extremely high strain rates. In each case, the overall stress response is dictated by the HE modelled equilibrium or intermediate springs [33]. The viscous damper in the Maxwell element doesn't contribute. Since HVE constitutively is a multiplicative integration of HE and viscoelastic (VE) models [30], thus, if the viscoelastic component diminishes then the overall HVE computed stress $\sigma_T$ also decreases. For linear viscoelastic model it is observed that stress relaxation function decays faster for higher spring stiffness '$K$' in each Maxwell component (linear spring-dashpot setup). Consequently, Relaxation time $\tau_R$ in $S(t,\varepsilon)$ decreases with increasing $K$ [35]. Thus, time dependent component of stress relaxation function diminishes exponentially by virtue of defined Prony-series viscoelastic material parameters and the resultant RMSD variations for varying $K$ configurations are obtained.

### 3.2.2 For varying number of connections:

For single-OL FEM using Abaqus Standard/Implicit SSD setup, effect of varying number of connections was analyzed for 2-nodes per axon (18-connections) and 5-nodes per axon (45 connections) cases, *cf.* **Figure 9**. Stiffness response results illustrated in **Figure 9** indicate stress relaxation and decrease in relative axonal stiffness in the case of increasing oligodendrocyte tethering. Stress-strain profile for 2-node per axon (18 connections) greater than 5-nodes per axon (45 connections) for given single-OL FEM. As observed in §3.2.1, with increasing $K$ value the stiffness decreased in the given FEM.

This trend is line with the observations and discussion presented in the §3.2.1. As number of tethering are increased in a model, more linear spring-dashpot elements get recruited for the total HVE Cauchy stress computation [15, 30]. With increasing viscoelastic elements, stress relaxation phenomenon also increases proportionally. While at first glance, there is no significant change noticed in stress profile with varying connections. But upon closer inspection, it is seen that **RMSD** for 18conn vs 45conn curves at *100K*: 0.000858 was higher than **RMSD** b/w 18conn vs 45conn curve at *10K*: 0.00011. Thus, depicting role of oligodendrocyte tethering in stress redistribution and axonal tissue softening when strain-rate dependent effects (viscoelastic parameters) are included in SSD analysis of harmonic/repetitive excitations at 50Hz.

### 3.3. Single-OL FEM – HVE Prony in Explicit Dynamics

In many TBI incidents the dynamic impact is applied for very short time period and lead to high non-linear deformations in the brain white matter. Hence, a 5-node per axon single-OL 3D FEM setup was analyzed for dynamic stretch for a short duration (step time, *t = 0.1s*) using the Explicit Dynamics (ED) step module in Abaqus 2020. For the ED model, the connection nodes on the axons and oligodendrocyte elements are defined a negligible mass to overcome any unconditionally unstable elements. The ED FEM model is evaluated keeping the same mesh settings as defined in the SSD FE model so that stress response can be compared.

Auto time-increment option in ED step chosen for the analysis, whereby the smallest mesh element dictates the stable time increment. In the ED FEM setup, total number of elements in 3D FEM : 409102. (3078 linear hexahedral elements of type C3D8R and 406024 linear tetrahedral elements of type C3D4). No Hybrid elements were defined for ED setup. For the hyper-elastic component value of $D_1$ was defined using $K_0$ from the Ogden strain density function constitutive relationships (Refer §2.2). In case of ED FEM, due to very fine mesh definition, large number of Abaqus solver determined stable time increments were needed to compute resultant stress response for each stretch case. Thus, ED proved very expensive and time-consuming FE setup for dynamic HVE response evaluation. In this paper, one representative ED case is evaluated and compared against SSD FEM setup for the same single-OL FEM configuration (45 connections at 10*K* oligodendrocyte spring stiffness). As evident from **Figure 10**, axons depicted high non-linear strain behavior at higher stretches and manifested stress accumulation characteristics over time when subjected to uniaxial dynamic tensile load. The amplified Cauchy-stress is governed by the HVE constitutive relation from **Equation 14**.

RMSD between the ED and SSD curves is calculated to be **0.182378** for 5-nodes per axon single-OL FEM at 10K. The non-linear stiffness response of the ED HVE-Prony model can be explained by referring to research on constitutive HVE models by *Ghahfarokhi et al.*[36] and *Hsu et al.*[37]. They report that at high strains, the relaxation time is a high-order function of induced strains. Since viscoelastic induced stress increases with the strain rate. Thus, computed overall HVE Cauchy-stress increases at higher strains and strain rates (greater stretch for the same time step duration of 0.1). The deviatoric component of the HVE Cauchy stress equation[37] drives the increase in axonal stiffness over strain history under dynamic load.

### 4. CONCLUSIONS

In the current study, a proposed 3D single-OL FEM framework was presented for to depict strain-rate and strain-history affect due to repetitive uniaxial stretch. An ensemble of simulations SSD and ED type FE cases for single-OL FEM configuration describing effect of oligodendrocytes tethering to axonal stiffness were investigated. These simulations were performed for brain matter properties defined in HE, HVE – frequency (50Hz) and HVE-time domain (Prony series) domains. For SSD setup, viscoelastic properties defined at 50Hz were chosen for analysis when subjected to external harmonic/repetitive load or concussions. For both HE and HVE axon material properties, 3D FEM numerical results indicated appearance of bending stresses along their tortuous path [7] inferring stress reversal due to inbuilt tortuosity. Magnitude of bending stresses or impending cerebral damage is dependent on axonal geometry; variation in brain mass; loading direction and frequency; and current state of the shear moduli (viscoelastic material definition in HVE).

For SSD simulations, HVE model depicted stress relaxation behavior in brain white matter when VE properties were analyzed in the frequency domain. Time-domain viscoelastic properties definition duly captured the stress increase due to increasing strains and strain rates. Prony Series parameters helped model the strain-rate affects in SSD FEM for HVE defined axons and ECM. Parametrization study of oligodendrocyte connections for the ensemble of cases in single-OL FEM indicated that increase in tethering aided in the stress relaxation. Thus, with increasing *K* value and number of connections, axonal stiffness decreased over the strain history. Such a behavior was noticed for both frequency and time domain viscoelastic model components in SSD simulations under pinning the importance of oligodendrocyte tethering on stress redistribution and tissue softening over repeated harmonic loads.

Over the years tissue level thresholds for axonal damage in CNS white matter have been extensively reported, but translation mechanics of macroscopic stresses and strain behaviors to microscopic scale is still not known. This study will pave way to understand micromechanical stress relaxation and strain history affects when HVE-modeled brain white matter is subjected to dynamic loads for short duration via discussed ED FEM. ED simulation results depicted role of viscoelastic component in improving axonal stiffness over specified strain history. Thus, oligodendrocyte connections are pivotal in improving stress-strain response to dynamic external impact loads such as blast, collisions, or concussions.

Proposed FEM for repetitive uniaxial tensile loading has potential limitations. First, the model approximated pure non-affine B.C. for entire stretch history, even though physiologically axons tend to exhibit more of transitional kinematics. FE models incorporating transition mechanism could yield high fidelity results [2]. Current hyper-viscoelastic micro-mechanical FEM approximates oligodendrocyte-axonal tethering by linear spring-dashpot connections. Moreover, in discussed hyper-elastic Ogden model, the impact of '$\alpha$' has not been explored and corresponding non-linearity due to it. In the HVE material model, linear viscoelastic model could be analyzed for inherent anisotropy often reported in viscoelastic brain soft tissues. Current results need further investigation by parameterizing it for all 5 connection configurations (shown in **Figure 3)** and also applying it on the multi-OL FEM submodels from our previous reported study[2, 7]. In terms of boundary condition for the SSD model, different amplitude curve functions and load frequencies (mild to severe TBI) could be evaluated to understand model stability for several ensemble cases. In this paper, only one representative single-OL FEM for explicit dynamic case could be presented. As a future objective, less computationally expensive meshing definitions and multiple load time steps definition can be explored to understand long term viscoelastic brain matter response. Explicit dynamics model for multiple time steps definition will help analyze damage initiation and strain accumulation effects in brain white matter.

Computationally feasible 3D FEM explicit dynamic models are key to understand damage initiation, high bending stresses related axon rupture and long-term axonal fatigue when subjected to either repeated harmonic or instantaneous dynamic impacts. Scope of proposed SSD and ED HVE brain matter models can be extended to evaluate structural response of aging or injured axons.

**ACKNOWLEDGEMENTS**

Support was provided by NSF Grants CMMI-1436743, CMMI-1437113, CMMI-1762774, CMMI-1763005.

**Table 1:** HE Material properties summary of SSD FE model

| Component | μ MPa | D 1/ MPa | α | Element Type |
|---|---|---|---|---|
| **Axon** | 2.15E-03 | 0 | 6.19 | C3D8H, C3D4H |
| **ECM** | 8.5 E-04 | 0 | 6.19 | C3D4H |

**Table 2:** Viscoelastic material properties - frequency domain [9]

| Component | Omega g* real | Omega g* imaginary | Omega k* real | Omega k* imaginary | Frequency |
|---|---|---|---|---|---|
| Axon | **0.6313** | **0.224439** | **0** | **0** | **0.05** |
| ECM | 0.6313 | 0.224438 | 0 | 0 | 0.05 |

**Table 3**: Prony series parameters used in VE material model [19]

| Component | $g_i$ Prony | $k_i$ Prony | $\tau_i$ Prony |
|---|---|---|---|
| **Axon** | 0.6039 | 0.60276 | 0.60097 |
| | 0.1083 | 0.10862 | 0.49866 |
| **ECM** | 0.50001 | 0.49995 | 0.00623 |
| | 0.25986 | 0.25745 | 0.90 |

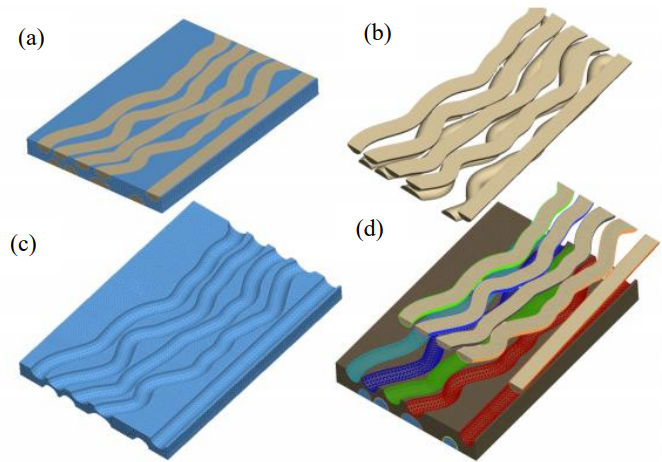

**FIGURE 1:** (a) FE Model of the ECM and axon assembly (b) FE model depicting the undulation of axons (c) FE model of ECM (d) Contact surfaces defining surface to surface contact between axons and ECM [7].

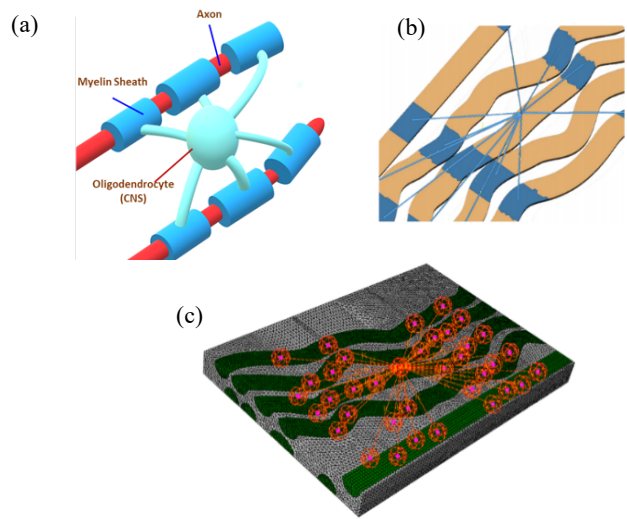

**FIGURE 2:** (a) Oligodendrocyte facilitating myelination in multiple axons in its vicinity. (b) A schematic representation of an oligodendrocyte tethering to axons at different locations via a sheath of myelin. (c) single-OL FE submodel: single oligodendrocyte tethering to surrounding axons[7].

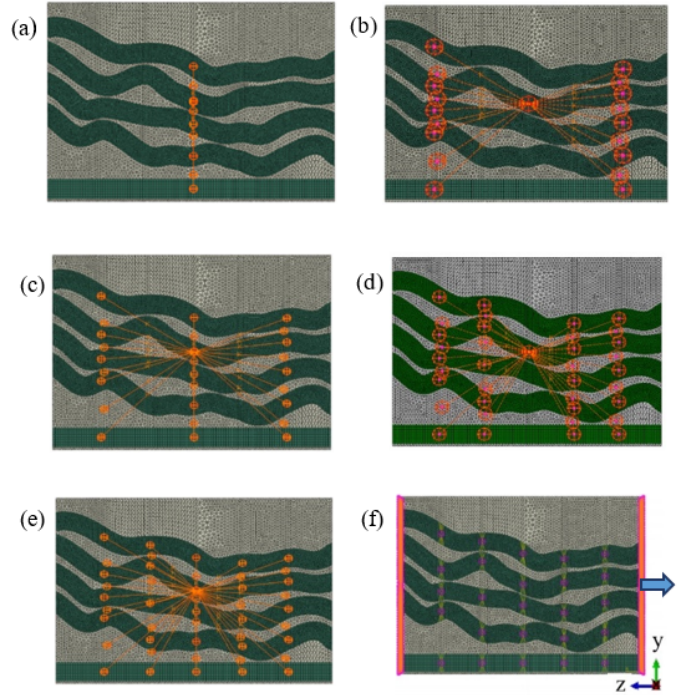

**FIGURE 3:** (a-e) Parameterization of number of connections between single-OL and each axon - Showing 1, 2, 3, 4 and 5 connections per axon, (f) boundary conditions for the FE model with the left end fixed and a stretch applied on the right [7].

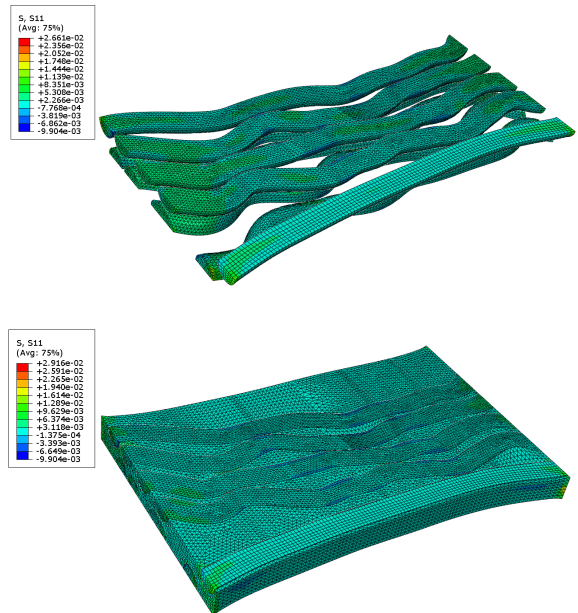

**FIGURE 4:** Von Mises stress contour for the Axons and the ECM at 20 percent applied stretch for 5-nodes per axon (single OL FEM – SSD FE setup) FE micro-mechanical model. Undulation of axons resulting in high stress in the concave regions. For 50Hz - SSD FE setup, stress is uniformly distributed between axons (top) and ECM (bottom).

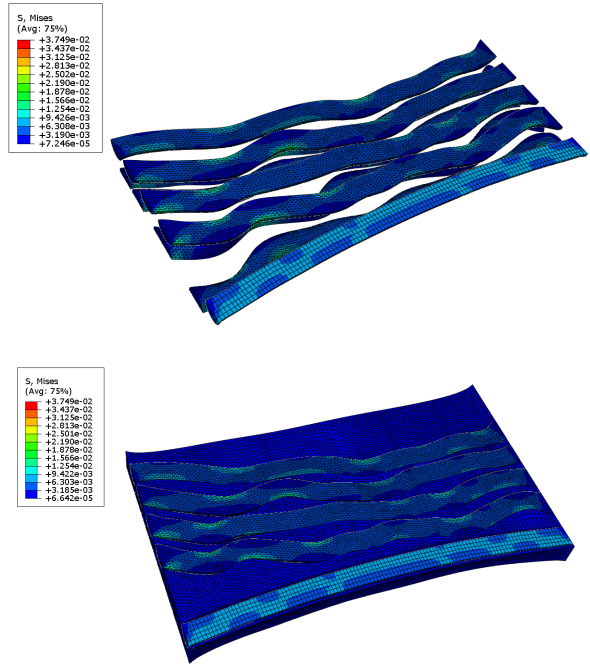

**FIGURE 5:** Von Mises stress contour for the axons and the ECM at 20 percent applied stretch for 5-nodes per axon (single OL - FEM) explicit dynamic (ED) - FE micro-mechanical model for single step time period of $t = 0.1$s. Undulation of axons led in high stresses in the concave regions as observed in pure HE models [7]. Bending stresses undergoing full reversal in axons from tension to compression. ED setup depicted non-linear strain history characteristics in HVE model. Bending stresses in axons over stretch history increases risk of fatigue failure. Since HVE -ED model captures non-linear strain history characteristics, von-Misses stress values are observed to be approximately 62% higher ED FEM compared to SSD FE setup for same configuration (see **Figure 5**). ED FEM results does account cumulative strains over time. Thus, exhaustive FE setup could help setting up fatigue damage model in future [2].

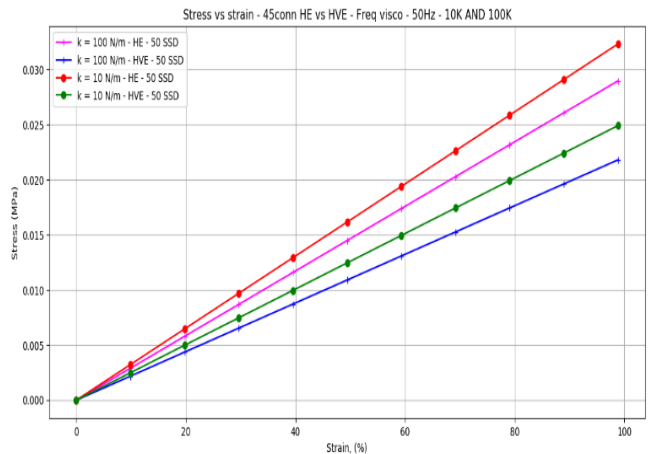

**FIGURE 6:** Stress ($\sigma$) versus stretch plot multi-oligo (single-OL FEM) FE model plotted for varying spring-dashpot stiffness ($K$). Impact of spring-dashpot stiffness ($K$) parameterization evaluated by overlaying 45 connections (5-nodes per axon, single-OL SSD results) and subsequent RMSD analysis done on the results to quantify relative decrease in stiffness for HVE model vs HE due to stress relaxation.

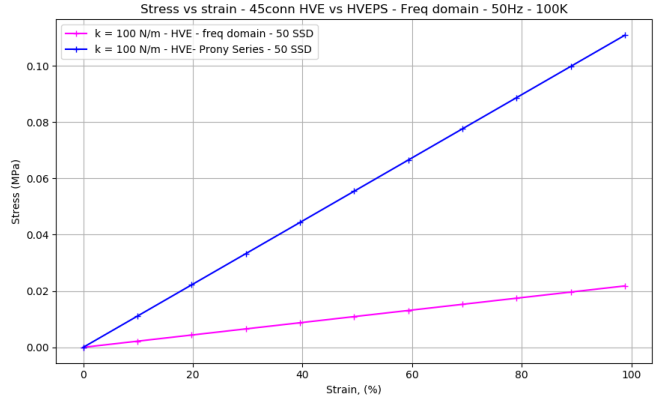

**FIGURE 7:** Stress-strain response for single-OL FEM (single-oligodendrocyte) SSD model. Simulations performed for 5 oligodendrocyte connections per axon (45 connections) for $K$ =100 N/m for HVE (frequency) and HVE-PS (Prony series) material definition. HVEPS showed greater stiffness over HVE-frequency model. Due to amplification in time-dependent component of the stress relaxation function. RMSD b/w curves at 100K: 0.05268.

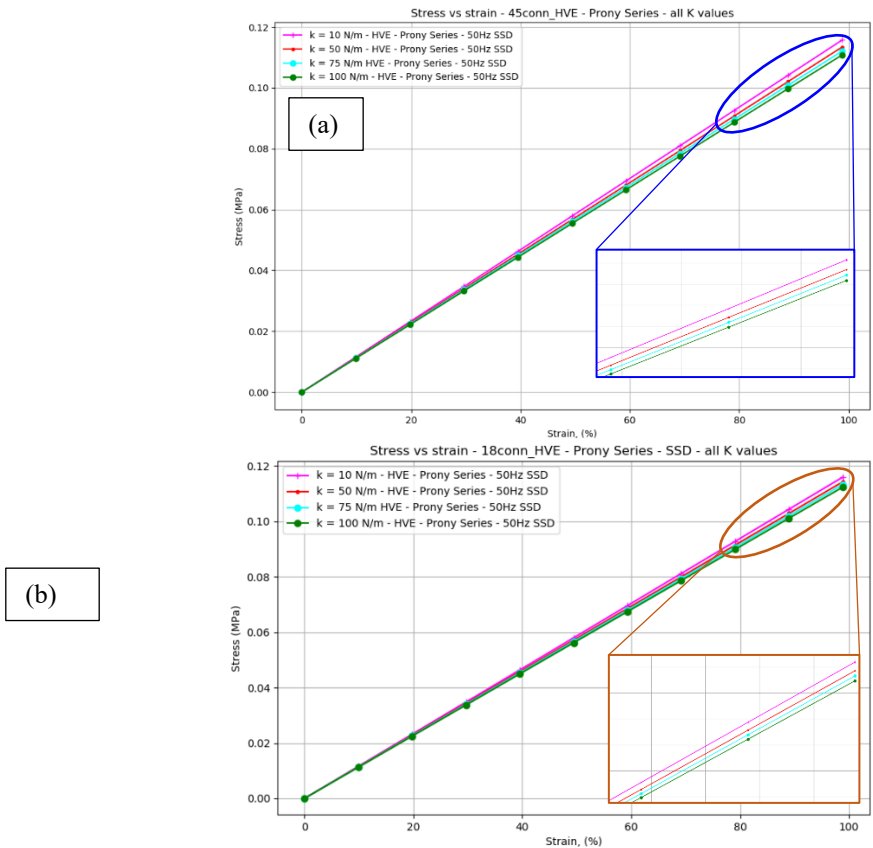

**FIGURE 8:** Stress-strain response for single-OL FEM SSD model. Simulations performed for (a) 5 oligodendrocyte connections per axon (45 connections) and (b) 18 connections models for varying $K$ = 10, 50, 75 and 100 N/m. HVE-PS (Prony Series) material definition used. Decrease in stiffness with increasing $K$ value in both ensemble.

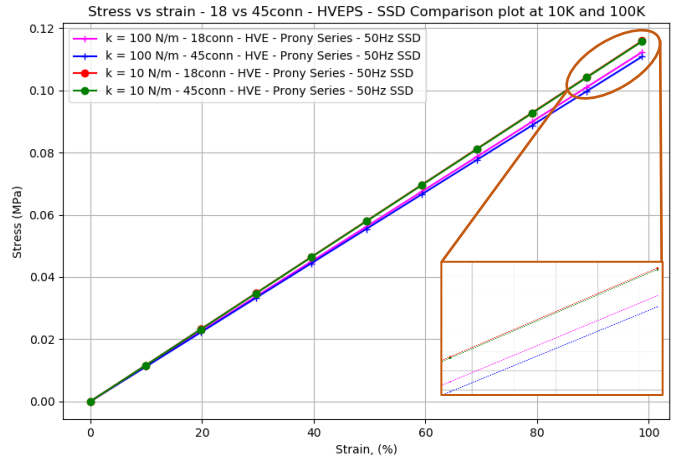

**FIGURE 9:** Stress-strain response for single-OL FEM SSD model. Simulations performed for 5 nodes per axon (45 connections) and 18 connections (2 nodes per axon) models at $K = 10$ and $100$ N/m. Axonal stiffness relaxation noticed for increasing connections. Decrease in stiffness with increasing K value in also both seen for K values.

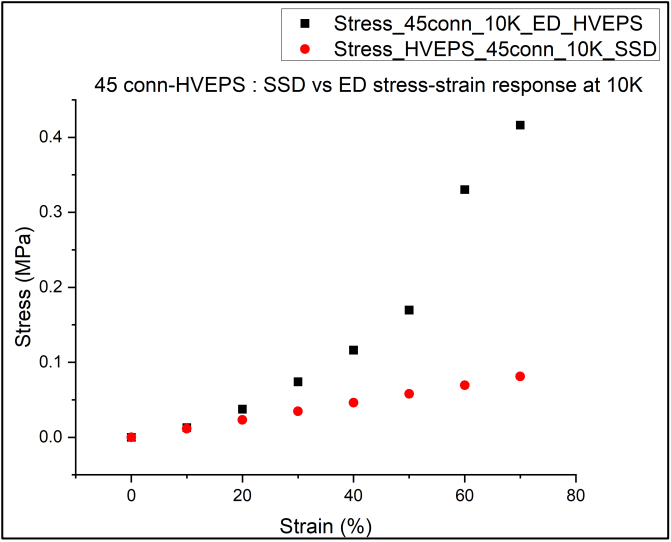

**FIGURE 10:** Stress-strain response for single-OL (45 connections) model at $K = 10$ for SSD FEM vs Explicit Dynamic (ED) setup to depict non-linear strain behavior in axons due to strain over time.